\documentclass[12pt]{emulateapj}

\usepackage{natbib}
\usepackage{graphicx}
\usepackage{amssymb}
\usepackage{epstopdf}
\DeclareMathSymbol{\varOmega}{\mathord}{letters}{"0A}
\DeclareMathSymbol{\varSigma}{\mathord}{letters}{"06}

\DeclareMathSymbol{\varPsi}{\mathord}{letters}{"09}

\newcommand{\Eq}[1]{equation\,(\ref{#1})}
\newcommand{\Eqs}[2]{equations (\ref{#1}) and~(\ref{#2})}

\newcommand{\Fig}[1]{Fig.~\ref{#1}}

\newcommand{\eqfrac}[2]{\left(\frac{#1}{#2}\right)}

 \newcommand{\Mdot}{\dot{M}}
\newcommand{\Msun}{M_{\odot}} 
\newcommand{\Lsun}{L_{\odot}}

\newcommand{\Mjup}{M_{\rm Jup}}
\newcommand{\lsim}{\mathrel{\rlap{\lower4pt\hbox{\hskip1pt$\sim$}}
    \raise1pt\hbox{$<$}}}                
\newcommand{\gsim}{\mathrel{\rlap{\lower4pt\hbox{\hskip1pt$\sim$}}
    \raise1pt\hbox{$>$}}}                
\begin{document}

\title{The Runts of the Litter: Why  planets formed through gravitational instability can only be failed binary stars}
\author{Kaitlin M. Kratter}
\affil{Department of Astronomy and Astrophysics, University of Toronto, 50 St. George Street, Toronto ON, M5S 3H4, Canada}
\author{Ruth A. Murray-Clay}
\affil{Harvard-Smithsonian Center for Astrophysics,  60 Garden Street, MS-51, Cambridge, MA 02138, USA}
\author{Andrew N. Youdin}
\affil{ Canadian Institute for Theoretical Astrophysics, University of Toronto,  60 St. George Street, Toronto, ON, M5S 3H8, Canada}

\begin{abstract}
Recent direct imaging discoveries suggest a new class of  massive, distant planets 
around A stars. These widely separated giants have been interpreted as signs of 
planet formation driven by gravitational instability, but the viability of this 
mechanism is not clear cut.  In this paper, we first discuss the local requirements 
for fragmentation and the initial fragment mass scales. We then consider whether the 
fragment's subsequent growth can be terminated within the planetary mass regime. 
 Finally, we place disks in the larger context of star formation and disk evolution models.
 We find that in order for gravitational instability to produce  planets, disks must be atypically
 cold in order to reduce the initial fragment mass.
In addition, fragmentation must occur during a narrow window of disk evolution, 
after infall has mostly ceased, but while the disk is still sufficiently massive to undergo 
gravitational instability. Under more  typical conditions, disk-born objects will likely
 grow well above the deuterium burning planetary mass limit.  
We conclude that if planets are formed by gravitational instability, they must be the low mass tail 
of the distribution of disk-born companions. To validate this theory, on-going direct 
imaging surveys must find a greater abundance of brown dwarf and M-star 
companions to A-stars.  Their absence would suggest planet formation by a different mechanism such as 
core accretion, which is consistent with the debris disks detected in these systems. 
\end{abstract}
\keywords{planet formation, accretion disks, binaries}

\maketitle
\section{Introduction}
Motivated by the recent discovery of massive planets on wide orbits, we  explore 
the requirements for making gas giants at large separations from their host star via 
gravitational instability, hereafter, GI. In particular, we consider the formation 
mechanism for the system HR 8799 which contains three $\sim$$10\Mjup$ objects 
orbiting at distances between $\sim$ 30 and 70 AU \citep{Marois08}. The 
standard core accretion model for planet formation, already strained in the outer 
solar system, has difficulty explaining the presence of these objects.  While GI is an
unlikely formation mechanism for close in planets 
\citep{2005ApJ...621L..69R}, for more widely separated planets, or sub-stellar companions, 
 the viability GI-driven fragmentation deserves further investigation.

In the inner regions of a protoplanetary disk, gas cannot cool quickly enough to 
allow a gravitationally unstable disk to fragment into planets 
\citep{2005ApJ...621L..69R,ML2005}.  For this reason, core accretion---in which 
solid planetesimals collide and grow into a massive core which then accretes a 
gaseous envelope---has emerged as the preferred mechanism for forming planets 
at stellar separations $\lesssim 10$ AU.  Planets at wider separations
have only recently been discovered by direct imaging around the A-stars HR 8799 
\citep{Marois08}, Fomalhaut \citep{2008Sci...322.1345K}, and possibly 
Beta Pic \citep{2009A&A...493L..21L}. Searches at large radii surrounding solar-type stars have yet to turn up similar companions \citep{2009arXiv0909.4531N}. Standard core accretion models cannot 
form these planets, though further investigation is warranted.

In favor of this possibility, all three systems show some evidence of processes 
related to core accretion: all have infrared excess due to massive debris disks at 
large radii. This is at least partially a selection effect as these systems were 
targeted due to the disks' presence.  Nevertheless, these debris disks are 
composed of reprocessed grains from collisions of planetesimals. The disks' long 
lifetimes prohibit a primordial origin for small grains---they are removed quickly
by  radiation pressure and  Poynting-Roberston drag \citep{1984ApJ...278L..23A} and so must be 
regenerated from collisions between larger bodies that formed through the 
coagulation of solids at early times. Therefore, planetesimal formation, a necessary 
ingredient in core accretion models, has taken place  \citep[e.g.,][]{YS02,ChiangYoudrev09}.

In addition, other A-stars host planets at 1--2 AU \citep{2007ApJ...665..785J}
 which, although they have a distinct semi-major axis distribution from 
 planets orbiting G and M stars, are likely formed by core accretion.  If future surveys demonstrate 
 that this distribution extends smoothly to wide separation planets, then simplicity
 would argue against a distinct formation mechanism for the wide giants.

Yet the standard core accretion model faces a serious problem at large distances.  
The observed lifetimes of gas disks are short, at most a few Myr 
\citep{1992ApJ...397..613H,2006ApJ...648.1206J}. In contrast, typical core accretion times 
increase with radius and exceed 10 Myr beyond 20 AU  \citep[e.g.][]{2001Icar..153..224L,2004ARA&A..42..549G}.
Whether or not this theoretical difficulty can be overcome will require careful 
modeling of the interactions between planetesimals and the young gas disk. 

Could wide orbit planets have formed at smaller radii, and migrated outwards? 
\cite{Dodson09}, have investigated the possibility of forming the HR 8799 system 
via scattering in the absence of dynamically important gas, but find that putting 
three massive planets into such closely spaced yet wide orbits is unlikely.  
\cite{crida09} have suggested that under favorable circumstances 
outward migration in resonances might be feasible.  
Alternatively, the core of a giant planet could be scattered outward by a planet, 
or migrate outward before accreting its gas envelope, either by interactions with
the gas disk (Type III migration; e.g. \citealt{2003ApJ...588..494M}) or with planetesimals
embedded in the gas (Capobianco, Duncan, \& Levison, in prep). 
Neither mechanism has yet been shown to move a core to such large distances,
though this possibility has not been ruled out.

Given these difficulties, it is natural to search for other formation mechanisms, and 
GI \citep{1997Sci...276.1836B} stands out as a promising alternative.  If any planets form
 by GI, the recently discovered directly imaged planets are the most likely candidates 
\citep{ RR09,2009ApJ...695L..53B,2009ApJ...702L.163N}.  In this paper, 
we examine this possibility in more detail, considering the expected mass scale of fragments
and the effect of global disk evolution on the formation process.

The inferred masses for the HR 8799 planets are close to the deuterium 
burning limit of $13\Mjup$ \citep{chabrier2000rev}.  For simplicity, we take this
as a the dividing line between planets and brown dwarfs and we refer to the HR 8799 objects
as planets throughout.  However, there is no reason for a given formation mechanism to 
function only above or below this threshold, and in fact, we will argue that if the HR 8799 planets formed by
GI, their histories are more akin to those of higher-mass brown dwarfs than to lower-mass planets.

To constrain GI as a mechanism for wide giant planet formation, we set the stage by 
describing the HR 8799 system in \S \ref{sec-HR8799}. We review the standard
requirements for fragmentation  in \S\ref{idealGI} and discuss the initial mass scale 
of fragments in \S\ref{sec-masses}.  In \S\ref{sec-GIaccrete} we show that under typical disk 
conditions, fragments will continue to accrete to higher masses. We then discuss important 
global constraints on planet formation provided by star formation models, disk 
evolution timescales, and migration mechanisms in \S \ref{sec-starform}, and \S \ref{sec-migration}.
 We compare predictions of our analysis with the known wide substellar companions and 
 exoplanets in \S \ref{sec-observe}, suggesting that future observations will provide a 
 definitive answer to the formation mechanism for HR 8799. 
In the appendices we  re-examine the heating and cooling properties of disks that 
are passively and actively heated, with special attention to the implications for 
irradiated disks, which become increasingly relevant for more massive stars.
 
  \section{The HR 8799 system}\label{sec-HR8799}
 The planets around HR 8799 probe a previously unexplored region of 
 parameter space \citep{Marois08,2009ApJ...694L.148L} because they are 
 more distant from their host star. The three companions to HR 8799 are observed
 at separations from their host star  of 24, 38, and 68 AU. Their masses, estimated 
 using the observed luminosities of  the planets in conjunction with cooling models,
  have nominal values of 10, 10, and 7 $\Mjup$, respectively.  A range in total mass 
  of 19--37 $\Mjup$ is derived from uncertainties in the age of the host star  \citep{Marois08}.  
Interpretation of the cooling models generates substantial additional systematic 
uncertainty---recent measurements suggest that these models may overpredict the 
masses of brown dwarfs by $\sim$25\% \citep{2009ApJ...692..729D}.
 \cite{FabMur09} have demonstrated that for planetary masses in the stated range, 
 orbital stability over the age of the system requires that the planets occupy at least one mean-motion resonance, and that for doubly-resonant orbital configurations total masses of up to at 
 least $54\Mjup$ can be stable.  
 
HR 8799 has been called a ``scaled-up solar system" in terms of the stellar flux 
incident on its giant planets \citep{Marois08,2009ApJ...694L.148L}. 
However for understanding the formation of this system it is more useful to consider 
dynamical times, and disk mass requirements.  Because the dynamical time at fixed radius scales only as $M_*^{1/2}$, the dynamical times are larger at the locations of the HR 8799 planets that at the solar system giants.
 Since the total mass in planets greatly exceeds the $\sim 1.5~ \Mjup$ in the solar system, 
 we can infer that (as with some other extrasolar systems) the primordial disk around
 HR 8799 was more massive than the solar nebula and/or there was greater efficiency 
 of planet formation, especially in the retention of gas.
Compared to solar system giants, longer dynamical times make core 
accretion more difficult and larger disk masses make GI more plausible.

\section{Ideal Conditions for GI-driven Fragment Formation}\label{idealGI}
We first determine where, and under what local disk conditions, fragmentation by 
GI is possible.  Following \cite{Gam2001}, \cite{ML2005} and \cite{2005ApJ...621L..69R},
 we argue that for a disk with surface density $\Sigma$ 
and temperature $T$ to fragment, it must satisfy two criteria.  First, it must have 
enough self-gravity to counteract the stabilizing forces of gas pressure and 
rotational shear, as quantified by Toomre's Q:
\begin{equation}  \label{eq-Q}
Q  \equiv \frac{c_s \Omega}{\pi G \Sigma}  <  Q_o \sim 1 
\end{equation}
\citep{Saf1960, Toom1964}, where $c_s = \sqrt{kT/\mu}$ is the isothermal sound 
speed of the gas with mean particle weight $\mu = 2.3 m_{\rm H}$ appropriate 
for a molecular gas, $G$ is the gravitational constant, $k$ is the Boltzmann constant, 
and $\Omega$ is the orbital frequency. 
Equation (\ref{eq-Q}) specifies the onset of axisymmetric instabilities in linear 
theory that can give rise to bound clumps \citep{GLB1965}.\footnote{In a realistic 
disk model, clumps likely form within spiral arms formed via non-axisymmetric, 
non-linear instabilities,  although the critical value of Q at which fragmentation 
occurs should remain similar. }

The second criterion that must be satisfied for fragmentation to proceed is the so-called
cooling time criterion. The heat generated by the release of gravitational binding energy during 
the contraction of the fragment must be radiated away on the orbital timescale so 
that increased gas pressure does not stall further collapse \citep{Gam2001}.  This 
implies
\begin{equation} \label{eq-tcool}
t_{\rm cool} = {3 \gamma\Sigma c_s^2\over 32(\gamma - 1)}{f(\tau) \over \sigma T^4 }  \lesssim  \zeta \Omega^{-1}.
\end{equation}  
Here $\zeta$ is a constant of order unity, $\gamma$ is the adiabatic index of the gas, and $\sigma$ is the Stefan-Boltzmann 
constant.  We take $f(\tau) = 1/\tau + \tau$ \citep{2005ApJ...621L..69R}
for disk vertical optical depth $\tau = \kappa \Sigma/2$ and gas opacity $\kappa$.  
Numerical models of collapse in barotropic disks 
measure the critical value $\zeta$ through the inclusion of a loss term $u/t_{\rm cool}$ in the equation 
for the internal energy, $u$.  
Estimates of $\zeta$ range from $\sim 3-12$, depending on  $\gamma$ \citep{Rice05}, the numerical 
implementation of cooling \citep{Clarke07}, and the vertical stratification in the disk.
We assume $\gamma = 7/5$, appropriate for molecular hydrogen, and we adopt $\zeta = 3$ here.
Although $\zeta$ was measured in disks whose temperature is controlled by viscous 
heating, we show in Appendix \ref{sec-irradcool} that the same expression (modulo 
slightly different coefficients) should apply when irradiation sets the disk 
temperature, as will be the case in disks prone to fragmentation (see Appendix 
\ref{sec-visc}). 

A disk satisfying Toomre's criterion for instability (equation \ref{eq-Q}) but not the cooling time criterion (equation \ref{eq-tcool}) experiences GI-driven angular momentum transport which regulates the surface density of the disk so that $Q \sim Q_o\sim 1$ and $Q$ does not reach substantially smaller values (c.f. Appendix \ref{sec-irradcool}).  We can therefore use Toomre's criterion to define a relationship between $\Sigma$ and $T$ at fragmentation, as a function of period:
\begin{equation}\label{eq-sigmaq}
\Sigma  =   \frac{c_s \Omega}{\pi G Q_o}  = f_q \sqrt{T}\Omega 
\end{equation}
where for convenience we define $f_q \equiv (k/\mu)^{1/2}(\pi G Q_o)^{-1}$.
We shall hereafter set $Q_o = 1$.

Given \Eq{eq-sigmaq}, we can rewrite \Eq{eq-tcool} to generate a single criterion for fragmentation
which depends on temperature and location:
\begin{eqnarray}\label{eq-tcoolB}
\frac{\Omega t_{\rm cool}}{\zeta} & = & (f_q f_t)~ \frac{\Omega^{2}}{T^{5/2}}f(\tau)\leq1,  
\end{eqnarray}
where $f_t \equiv (3/32)\gamma(\gamma-1)^{-1}k(\mu\sigma\zeta)^{-1}$.
Somewhat counterintuitively, the critical cooling constraint requires that a disk be sufficiently hot
to fragment. The value of $f(\tau)$ depends on both $\Omega$ and $T$.  Evaluating this criterion
 relies on the disk opacity, which we return to in \S\ref{sec-opacity}.

\section{Minimum fragment masses and separations}\label{sec-masses} 
\subsection{Initial masses of GI-born fragments}
We take the initial mass of a fragment to be the mass enclosed within the radius of 
the most unstable wavelength, $\lambda_Q = 2\pi H$ in a $Q=1$ disk, or:
\begin{equation} 
M_{\rm frag} \approx\varSigma (2 \pi H)^2 
\end{equation}
\citep{2007MNRAS.374..515L}, where $H  = c_s/\Omega$ is the disk scaleheight. 
 \cite{Cossins09} have shown that even when the GI is non-axisymmetric, 
the most unstable axisymmetric wavelength $\lambda_Q$ is one of the 
dominant growing modes, suggesting that this is a reasonable estimate.
While more numerical follow up will be necessary to pin down the true distribution 
of fragments born through GI, at present simulations 
show that this estimate may well be a lower limit, but is the correct order of 
magnitude \citep{2009ApJ...695L..53B, 2009MNRAS.392..413S}.

Using \Eq{eq-sigmaq}, we rewrite the fragment mass explicitly as a function of temperature
and location:
\begin{eqnarray}\label{eq-mfragt}
M_{\rm frag}  \approx 4\pi \eqfrac{H}{r}^3 M_* = f_m \frac{T^{3/2}}{\Omega} 
\end{eqnarray}
where $f_m \equiv (2\pi)^2 f_q k/\mu$.
Equation (\ref{eq-mfragt}) demonstrates that at a given disk location, fragment 
masses depend only on temperature, with lower temperatures generating smaller 
fragments, subject  to the minimum temperature required for fragmentation by \Eq{eq-tcoolB}.

\cite{2005ApJ...621L..69R} pointed out that there exists an absolute minimum
 fragment mass at any disk location, when the disk satisfies  the equalities in \Eq{eq-Q}
 and \Eq{eq-tcool} and has $\tau=1$.  The minimum temperature required for fragmentation 
 scales as $T \propto (\tau+1/\tau)^{2/5}$ (equation \ref{eq-tcoolB}),  so  the critical temperature
 and fragment mass are minimized at  $\tau = 1$, the optical depth for which cooling is most efficient.
The corresponding minimum mass as a function of location is:
\begin{eqnarray} \label{eq-minmass}
M_{\rm f,min} &=& f_m (f_q f_t)^{3/5} \Omega^{1/5} \\ 
&=& 1.5\Mjup \eqfrac{r}{100\rm{~AU}}^{-3/10}\eqfrac{M_*}{1.5\Msun}^{1/10}
\end{eqnarray}
which occurs for disk temperatures:
\begin{equation} \label{eq-mintemp}
T_{\rm f,min} = 7 \rm{K} \eqfrac{r}{100\rm{~AU}}^{-6/5}\eqfrac{M_*}{1.5\Msun}^{2/5}
 \end{equation} 
Equation (\ref{eq-minmass}) corresponds to $Q=1$, $\Omega t_{\rm cool} = 3$, and $\tau = 1$. 
This minimum mass is only achieved for temperatures consistent with $T_{\rm f,min}$. Once an
 opacity law is specified which relates $T$ and $\tau$, the problem becomes overconstrained---
these three criteria can only be satisfied at a single disk location, and equations 
\ref{eq-minmass}-\ref{eq-mintemp} are valid at only one radius in the disk. 
We now proceed to evaluate the critical disk temperatures and fragment masses for
realistic opacity laws,  demonstrating that planetary-mass fragments can
only form at large separations from their host star. 

\subsection{Opacity}\label{sec-opacity}
At low temperatures, when $T \lesssim 155$ K, ice grains are the dominant source 
of opacity \citep{Pollack85}. Above $\sim$$155$ K, ices begin to  sublimate.
In the cold regime applicable to the outer regions of protoplanetary disks, we consider
two realistic opacity laws, one which is characteristic of grains in the interstellar medium (ISM), and one which is characteristic
of grains that have grown to larger sizes due to processes within the disk. We assume a Rosseland 
mean opacity which scales as 
\begin{equation} \label{eq-kappa}
\kappa \approx \kappa_\beta T^\beta
\end{equation}
where the both the exponent, $\beta$, and the constant $\kappa_\beta$ are not well constrained in
protoplanetary disks. They depend on 
the number-size distribution and composition of the dust grains, and on the dust-to-gas ratio.  
For ISM like grains, the Rosseland mean opacity may be approximated by
an opacity law with $\beta=2$, or
\begin{equation}
 \kappa \approx \kappa_2 T^2
\end{equation}
\citep{1996Icar..124...62P,1994ApJ...427..987B, 2003A&A...410..611S} as long as 
the ice grains dominating the opacity are smaller than a few tens of 
microns.  This opacity law is observationally confirmed in the ISM \citep[and 
references therein]{2000prpl.conf..533B}. 
 For our fiducial model we use $\kappa_2 \approx 5 \times 10^{-4} {\rm cm^2/g}$  for $T$ in Kelvin,
 a fit to the standard dust model by \cite{2003A&A...410..611S}. 
Throughout, we quote $\kappa$ per gram of gas for a dust-to-gas ratio of $10^{-2}$.

As grains grow larger,  they eventually exceed the wavelength of the incident radiation, and so
the opacity is determined by the geometric optics limit.  In this case the Rosseland mean opacity is
independent of temperature, so the exponent $\beta=0$. In this limit:
\begin{equation} \label{eq-ind_kap}
\kappa \approx \kappa_0 
\end{equation}
For a fixed dust mass in grains of size $s$, $\kappa_0 \propto s^{-1}$.  For our fiducial model we choose 
$\kappa_0 \approx 0.24~\rm{cm^2/g}$ which is valid for $T \gtrsim 20$ K and typical
 grain sizes of order 300$\mu m $  \citep[see Figure 6 of][]{Pollack85}. 
  
Observations of  emission from optically thin protoplanetary disks show 
evidence of grain growth at millimeter wavelengths.  Specifically the measured 
opacities $\kappa_\nu \propto \nu^\alpha$ with $\alpha \simeq 0.5$---$1.5$ in the 
Rayleigh-Jeans tail imply particle growth toward the millimeter wavelengths of the 
observations \citep{2001ApJ...553..321D}.  Although most observed disks have had more time for grain 
growth to proceed,  Class 0  sources also show evidence of grain growth \citep{2001ApJ...553..321D}.
 Alternatively, these objects could be optically 
thick at millimeter wavelengths, which would mimic the effects of grain growth.  

Using these opacity laws we see that cooling proceeds with a different functional
 form in the optically thick and optically 
thin limits. In the optically thick regime, $f(\tau) \approx \tau = (\kappa_\beta T^{\beta} \Sigma)/2 $,
which when combined with \Eq{eq-tcoolB} indicates that to fragment, the disk must have temperature
\begin{equation}\label{eq-tcoolcritthick}
T > \left(f_t f_q^2/2\right)^{1/({2-\beta}) }\left(\Omega^3{\kappa_\beta}\right)^{1/({2-\beta})}
\end{equation}
for $\beta \ne 2$.  In the special case $\beta = 2$, the fragmentation constraint is temperature-independent, 
as discussed in \S\ref{sec-ism}.

In the optically thin regime, $f(\tau)= 1/\tau = 2/(\kappa_\beta T^{\beta} \Sigma)$.
The cooling time is independent of $\Sigma$, so we can rewrite \Eq{eq-tcoolB} as:
\begin{equation}\label{eq-tcoolcritthin}
T > (2 f_t)^{1/({3+\beta})}\left(\frac{\Omega}{\kappa_\beta }\right)^{1/({3+\beta})},
\end{equation}

\begin{figure}
\centering
\scalebox{0.5}{\includegraphics{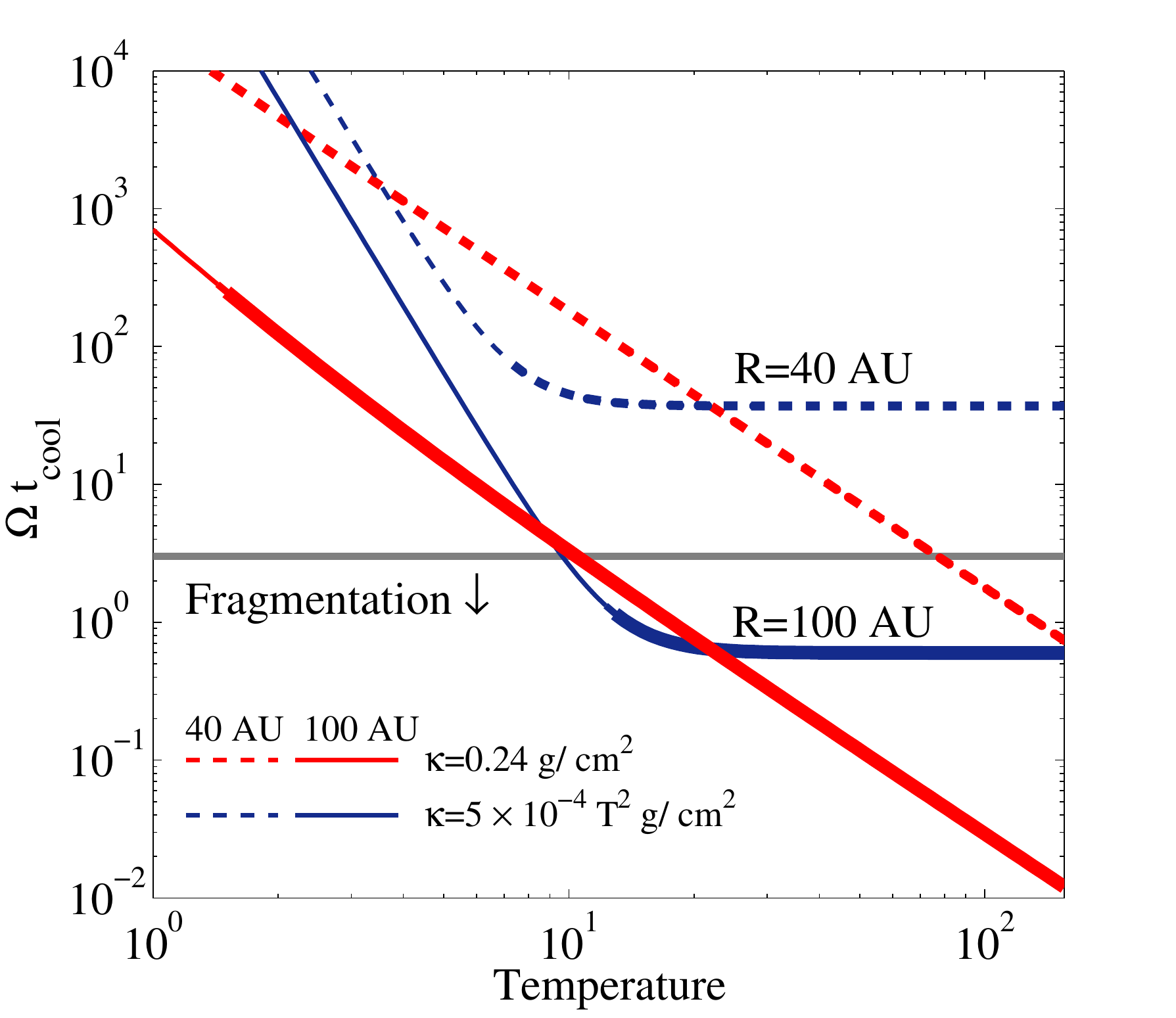}}
\caption{ The disk cooling time as a function of tempertature for different opacity laws 
at radii of 40 AU (dashed) and 100 AU (solid). The cooling time is calculated assuming that 
$Q=1$. The temperature independent (large grain) opacity law is shown in red, while the 
ISM opacity law: $\kappa \propto T^2$ is shown in blue. The line thickness indicates the
 optical  depth regime. When lines drop below the critical cooling time (grey), disk fragmentation
  can occur.  The bend in the ISM opacity curve indicates that in the optically thick regime, 
  the cooling time becomes constant as a function of temperature. } 
\label{fig-tcool}
\end{figure}

\Fig{fig-tcool} shows the dependence of the cooling time on disk temperature 
for each opacity law at two different radii. Since fragmentation can only occur when 
$\Omega t_{\rm cool} < \zeta$, the minimum temperatures at which fragmentation is allowed
 are specified by the intersection of the cooling curves with the $\Omega t_{\rm cool}$
boundary. 

\subsubsection{ Small grain opacity law}\label{sec-ism}
For an optically thin disk with $\beta = 2$, \Eq{eq-tcoolcritthin} implies that in order 
to fragment, the disk must have temperatures in excess of:
\begin{eqnarray} \label{eq-T2thin}
T &>& T_{\rm thin} = 9 \rm{K} \eqfrac{r}{100\rm{~AU}}^{-3/10}\\
& & \eqfrac{\kappa_2}{5\times 10^{-4} \rm{cm^2/g}}^{-1/5}\eqfrac{M_*}{1.5\Msun}^{1/10} \nonumber
 \end{eqnarray}
Colder disks, even when optically thin, cannot cool quickly enough to fragment.

\begin{figure}
\centering
\scalebox{.6}{\includegraphics{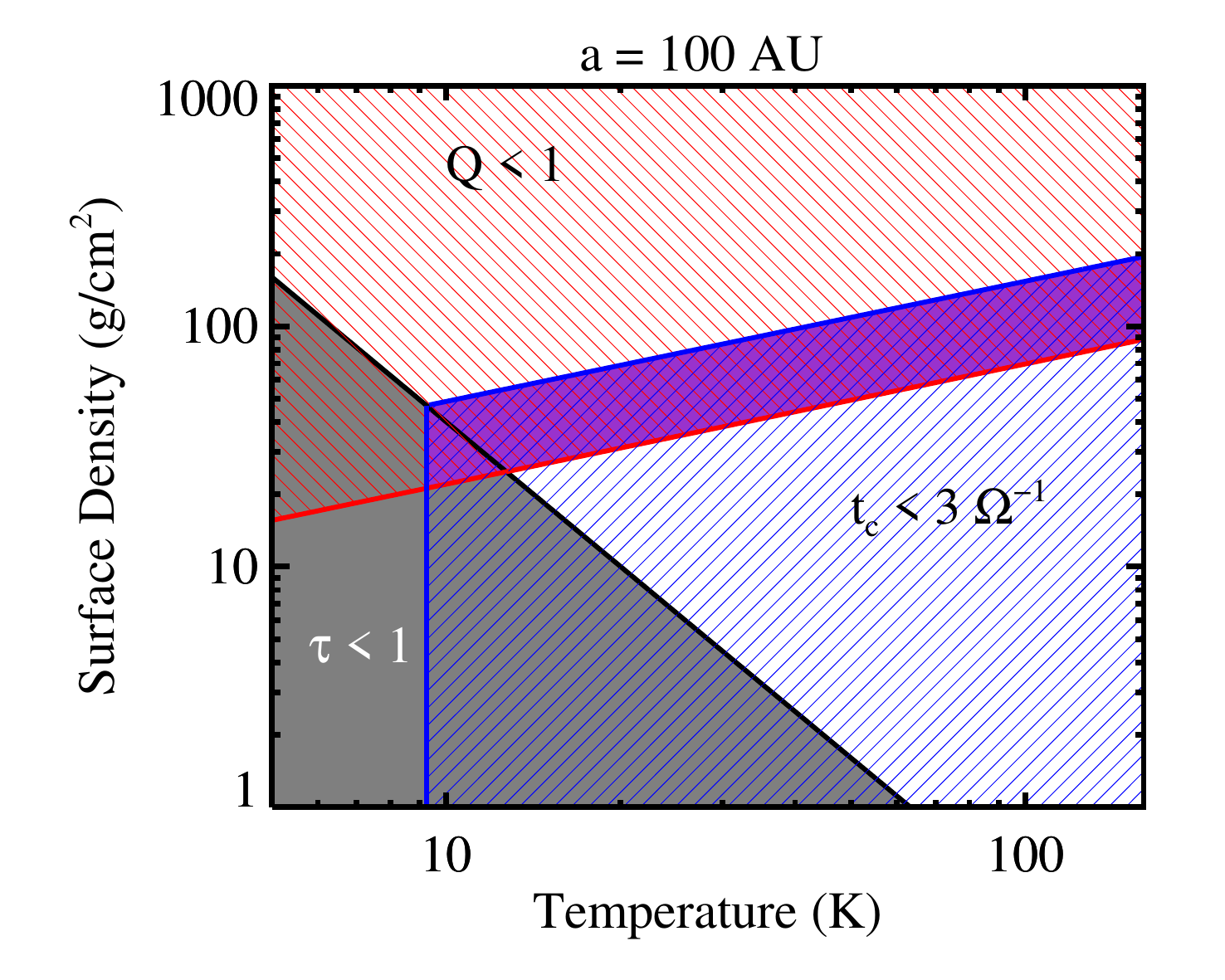}}
\caption{Fragmentation can only occur in the region of parameter space indicated by
 the overlapping hashed regions for ISM opacities at radii of 100 AU.  The upper, shaded 
 region (red) shows where Toomre's parameter $Q < 1$. The lower, shaded
region (blue) indicates where $t_{\rm cool} \le 3\Omega^{-1}$.   
At radii less than $70$ AU, fragmentation is prohibited because the two regions 
no longer overlap. That the boundaries of these regions are parallel lines reflects 
the $\kappa \propto T^2$ form of the ice-grain-dominated opacity at low temperatures.}
\label{fig-a100}
\end{figure}

For an optically thick disk, $\beta=2$ turns out to be a special case: the temperature dependence drops out of  \Eq{eq-tcoolcritthick}, giving instead a critical radius beyond which fragmentation can occur, independent of the disk temperature:
 \begin{equation} \label{eq-rminthick}
r \gtrsim 70 ~{\rm AU} \left(\frac{M_\ast}{1.5 M_\Sun}\right)^{1/3}\eqfrac{\kappa_2}{5\times 10^{-4}\rm{{cm^2}/{g}}}^{-2/9}.
\end{equation}
\cite{ML2005} first pointed out the existence of a minimum critical radius for fragmentation. 
In \Fig{fig-a100} we illustrate how the two fragmentation criteria create a radius rather than 
temperature cutoff. At radii larger than the critical radius defined above, any $\Sigma-T$ 
combination which gives $Q \leq 1$ will fragment (so long as the opacity law remains valid). 
At smaller radii, no combination of $\Sigma$ and $T$ which gives $Q \leq 1$ will fragment
 because the disk cannot simultaneously satisfy the cooling time criterion.  

\subsubsection{Large grain opacity law}
As shown in \Fig{fig-tcool}, for the large grain opacity, $\tau >1$ for all relevant temperatures. 
In this case \Eq{eq-tcoolcritthick} requires:
\begin{equation}\label{eq-tminz}
T >65 K \ \left(\frac{M_\ast}{1.5 M_\Sun}\right)^{3/4}\eqfrac{r}{43 \rm{~AU}}^{-9/4}
 \eqfrac{\kappa_0}{0.24\rm{~cm^2/g}}^{1/2}.
\end{equation}
The corresponding minimum mass for these temperatures is:
\begin{eqnarray}\label{eq-mminz}
M_{\rm min}&= & 13\Mjup  \eqfrac{M_*}{1.5\Msun}^{5/8}\eqfrac{r}{43\rm{AU}}^{-15/8} \\ \nonumber
&&\eqfrac{\kappa_0}{0.24 \rm{cm^2/g}}^{3/4}.
\end{eqnarray}
We have scaled \Eqs{eq-tminz}{eq-mminz} to an effective critical radius for this opacity law. 
Although fragmentation can occur inside 43 AU at sufficiently high temperatures, the fragments
exceed $13\Mjup$, making it irrelevant for planet formation. Smaller values of $\kappa_0$ 
move this boundary inward, allowing for fragmentation into lower mass objects at smaller radii, 
although the scaling with opacity is shallow. Grain growth to larger sizes could plausibly reduce 
 $\kappa_0$. If, for example, grains dominating the disk opacity have grown 
up to 1mm without altering the dust-to-gas ratio, \Eq{eq-ind_kap} implies that in the geometric
optics limit $\kappa_0 = 0.072 \rm{cm^2/g}$.   In this case, the minimum radius is pushed
inward to 26 AU \citep[see also][]{2009ApJ...702L.163N}.   

Thus far we have determined the minimum masses allowed as a function of radius with the 
temperature as a free parameter. We now calculate actual disk temperatures, which at large
 radii are typically higher than the minima. In this case we must evaluate fragment masses 
 using \Eq{eq-mfragt}.

 \subsection{Initial fragment masses with astrophysical disk temperatures}\label{sec-thermo}

To consider the case favorable to GI planet formation, we consider the lowest plausible disk 
temperatures in order to minimize the fragment masses. We estimate the disk temperature
 using the passive flared disk models of  \cite{1997ApJ...490..368C}. This model likely 
 underestimates the temperatures in actively accreting systems because it ignores
  significant ``backheating" from the infall envelope  and surrounding cloud \citep{1997ApJ...477..398C,ML2005}.  
Although viscous heating will also contribute to the temperature, we ignore its modest contribution
 to obtain the lowest reasonable temperatures and fragment masses. Disk irradiation dominates 
 over viscous heating in this regime (see Appendix B).

We consider the inner region where the disk is optically thick to blackbody 
radiation. In this regime, the temperature of a flared disk in radiative and hydrostatic 
equilibrium is:
\begin{equation} \label{eq-CG97tmid}
T_m = \left(\frac{\alpha_F}{4}\right)^{1/4}\left(\frac{R_*}{r}\right)^{1/2} T_* \propto L^{2/7}r^{-3/7}
\end{equation}
where $\alpha_F$ measures 
the grazing angle at which starlight hits the disk; $\alpha_F$ is dependent on the degree of disk flaring
measured at the  height of the photosphere \citep{1997ApJ...490..368C}. 
Grain settling may reduce the height of the photosphere (set here to 4 times the scaleheight)
and thus $\alpha_F$. For the standard radiative equilibrium model, the disk 
flaring scales approximately as $H/r \propto r^{2/7}$.  We shall find when we 
calculate the disk temperature that a gravitationally unstable disk remains
 optically thick, justifying the use of this formula.

To estimate the the stellar luminosity we use the stellar evolution models of 
\cite{krumholz07c}, which include both nuclear burning and accretion energy.
 The accretion luminosity depends on both the  current accretion rate and the accretion history ( in 
so far as it effects the stellar radius), so we obtain the lowest luminosity estimates 
by allowing the star to accrete at a constant, low accretion rate.  We use the stellar 
luminosity after accreting to 90\% of its current mass (or $1.35 \Msun$ assuming 
roughly 10\% is still in the disk).  We choose an accretion rate of  $10^{-7}\Msun/\rm{yr}$ 
as a lower bound because a star accretingat a lower accretion rate throughout its history 
has a formation timescale that is too long.  Accretion rates an order of magnitude larger
give comparable luminosity (when the star has only reached $1.35\Msun$) to the present
day luminosity of $5\Lsun$ \citep{Marois08}. Lowering the accretion rate below this 
value does not significantly lower the stellar luminosity because the accretion energy
contribution is small.

The luminosity calculated for an accretion rate of  $10^{-7}\Msun/\rm{yr}$
translates to temperatures of:
\begin{equation}\label{eq-minT}
T \approx  40 {\;\rm K} \; \left(\frac{r}{70 {\;\rm AU}}\right)^{-3/7},
\end{equation}
which we shall use as our fiducial temperature profile.
In the outer regions of the disk where fragmentation is allowed, the disk temperatures are of order 
$30-50$K. These temperatures exceed the minimum threshold for fragmentation, and so the mass of fragments 
will be set by equation (\ref{eq-mfragt}). 

Other analytic and numerical models of stellar irradiation predict temperatures in 
agreement with or higher than our estimate. \citep{2006ApJ...646..275R,2009ApJ...703..131O}.  
Similarly, models of disks in Ophiuchus have similar temperatures for 1 Myr old 
stars of lower mass (and thus luminosity), implying that our model temperatures 
are low, though not unrealistic \citep{2009ApJ...700.1502A}. 
\begin{figure}
\centering
\scalebox{0.4}{\includegraphics{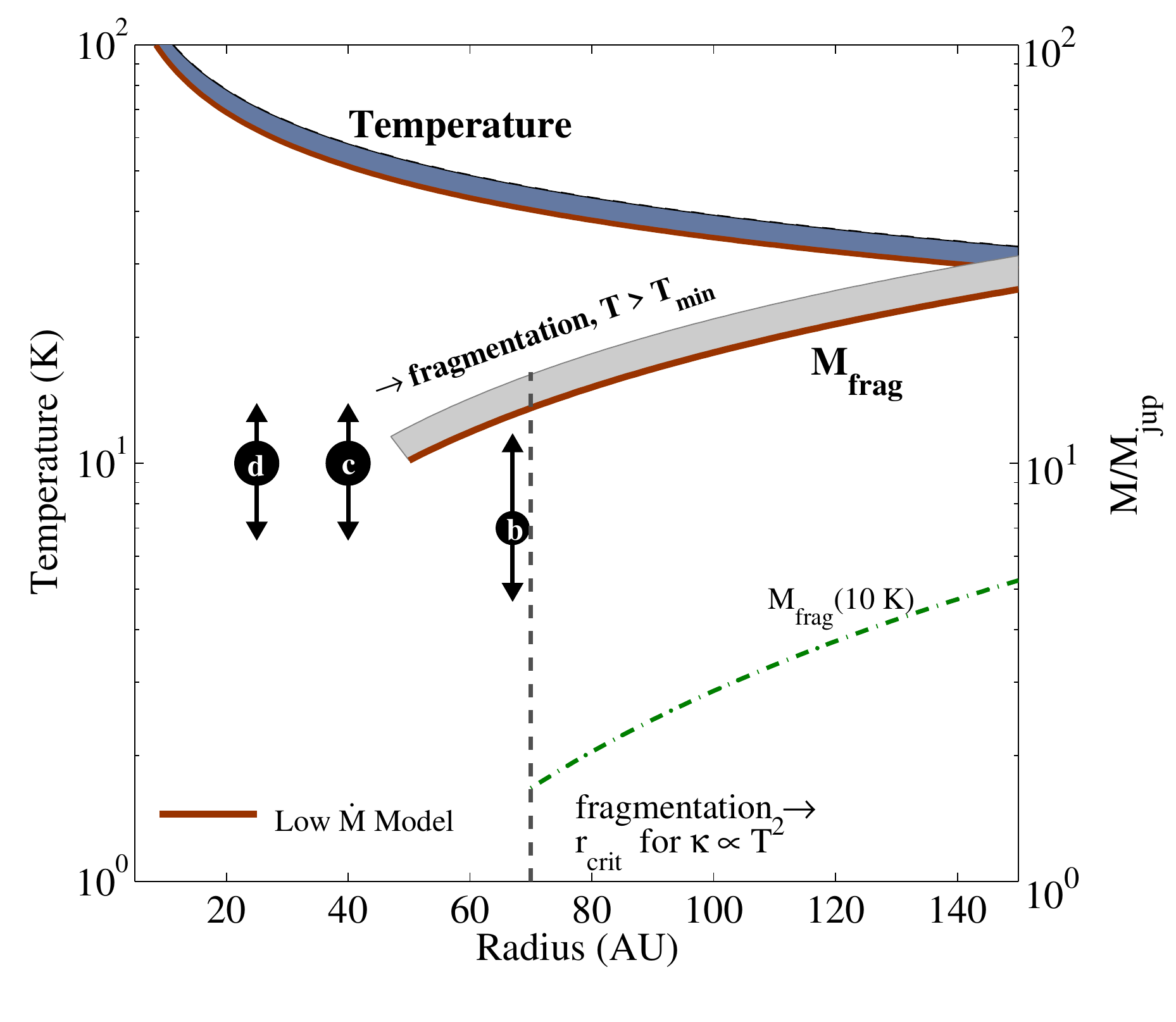}}
\caption{Depiction of the current configuration of HR 8799 and formation
 constraints for realistic disk temperatures. We show the lowest expected
 irradiated disk temperatures (blue) and  corresponding fragment masses (grey), 
 as a function of radius.  The lower bound on both regimes (burgundy)  is set
 by the irradiation model described in \S\ref{sec-thermo}, with $\Mdot = 10^{-7} \Msun\rm{/yr}$. 
The upper boundary is set by the current luminosity of HR 8799, $\sim 5 \Lsun$. 
The green dashed-dotted line shows the mass with disk temperatures of 10 K,
a lower limit provided by the cloud temperature. The vertical line shows the critical fragmentation
 radius for the ISM opacity law; fragmentation at smaller radii requires grain growth. 
 Fragment masses are shown for radii at which the irradiation temperatures 
 are high enough to satisfy the cooling time constraint of \Eq{eq-tminz}. At smaller radii,
 fragmentation is possible at higher disk temperatures, but the resulting fragments have 
 correspondingly higher masses, and planet formation is not possible.  }
\label{fig-cartoon}
\end{figure}

In \Fig{fig-cartoon} we illustrate the constraints on fragment masses from this 
irradiation model, calculated using \Eq{eq-mfragt}. We show the fiducial disk 
temperatures of \Eq{eq-minT}, along with temperatures consistent with
luminosities up to the present-day luminosity.  For our fiducial opacity laws, the 
expected fragment masses are only marginally consistent with GI planet formation---
fragments form near the upper mass limit of $13\Mjup$.  Lower opacities produced
 by grain growth might allow fragmentation into smaller objects at closer radii.
Whether grain-growth has proceeded to this extent in such young disks is unclear.

\section{Growth of fragments after formation}\label{sec-GIaccrete}
For realistic disk temperatures, it is conceivable that fragments may be born 
at several $\Mjup$. We now consider their subsequent growth, which may increase their 
expected mass by an order of magnitude or more.

The final mass of a planet depends sensitively on numerous disk properties 
(effective viscosity, column density, scaleheight) and the mass of the embedded 
object.  In order to constrain the mass to which a fragment will grow, we can 
compare it to two relevant mass scales: the disk isolation mass and the gap 
opening mass. 

Halting the growth of planetary mass objects is a relevant problem independent of the 
formation mechanism.  However the GI hypothesis requires that the disk is (or was 
recently) sufficiently massive to have $Q \sim 1$, implying that the disk is actively 
accreting.  The core accretion scenario does not face this restriction.

\subsection{Isolation Mass}\label{sec-isolation}
We estimate an upper mass limit for fragments by assuming that they accrete all 
of the matter within several Hill radii:
\begin{equation}\label{eq-Miso}
M_{\rm{iso}} \approx 4 \pi f_H\Sigma R_H r. 
\end{equation}
Here $f_H \sim 3.5$ is a numerical constant representing from how many Hill radii, 
$R_H = r (M_{\rm iso}/ 3M_*)^{1/3}$, the planet can accrete \citep{1987Icar...69..249L}. 

\begin{figure}
\centering
\scalebox{0.42}{\includegraphics{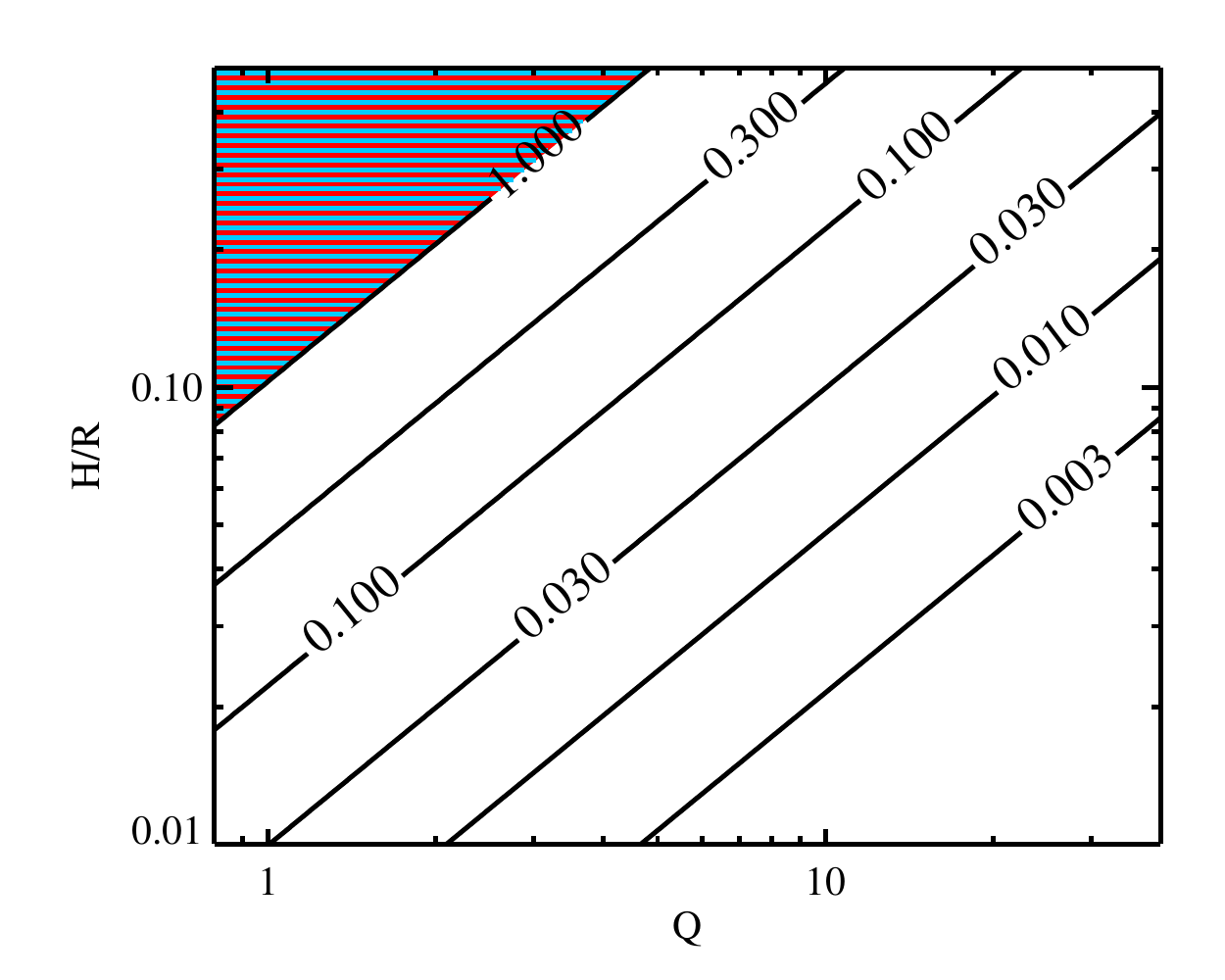}}
\caption{Contours of the ratio of planetary isolation mass to stellar mass as a 
function of  Toomre's $Q$ and the disk aspect ratio $H/r$, illustrating that the 
isolation mass is always large in unstable disks.  For disks with 
higher $Q$'s consistent with core accretion models, the isolation mass remains small.
The shaded region  indicates where the isolation mass exceeds the stellar mass. }
\label{allmasses}
\end{figure}

It is instructive to compare the ratio of the isolation mass to the stellar mass: 
\begin{equation} \label{eq-miso-m}
\frac{M_{\rm{iso}}}{M_*} = 4.6 f_H^{3/2}{Q^{-3/2}}\left(\frac{H}{r}\right)^{3/2}.
\end{equation}
We find that large isolation masses are always expected in gravitationally unstable 
disks. \Fig{allmasses}  illustrates the scaling of equation (\ref{eq-miso-m}) with $Q$ 
and the disk aspect ratio, $H/r$. For our fiducial temperature profile, $H/r \approx 0.09$ at 70 AU.
For low values of $Q$ and comparable $H/r$, the isolation mass exceeds 
 $10\%$ of the stellar mass. For the ideal disk values cited above (equation \ref{eq-rminthick}), 
 the isolation mass is:
 \begin{equation}
 M_{\rm{iso}} \approx  400\Mjup\left(\frac{ r}{ 70 ~\rm{AU}}\right)^{6/5} 
 \left(\frac{M_*}{1.5 \Msun}\right)^{1/10}. 
 \end{equation}
This  mass is nearly two orders of magnitude larger than the minimum mass.  
Growth beyond the isolation mass is possible either through mergers or 
introduction of fresh material to accrete by planet migration or disk spreading.
 
Objects which grow to isolation mass cannot be planets, and so we turn to 
mechanisms that truncate fragment growth below the isolation mass.

\subsection{Gap opening mass}\label{sec-gap}
Massive objects open gaps in their disks when gravitational torques are sufficiently 
strong to clear out nearby gas before viscous torques can replenish 
the region with new material. \citep{1986ApJ...309..846L,1999ApJ...514..344B}.  
The gap width is set by the balance between the two torques:
\begin{equation}\label{eq-gapwidth}
\frac{\Delta}{r} = \left(\frac{f_g q^2 }{3\pi \alpha}\frac{r^2}{H^2}\right)^{1/3},
\end{equation}
where $\Delta$ is the gap width, $f_g \approx 0.23$ is a geometric factor derived in 
\cite{1993prpl.conf..749L}, $q$ is the planet to star mass ratio, and $\alpha$ is the 
\cite{SS1973} effective viscosity. This can be used to derive the standard minimum 
gap opening mass by requiring that $\Delta > H$:

\begin{eqnarray}\label{eq-gapmass}
q& >& \left(\frac{H}{r}\right)^{5/2}\sqrt{\frac{3\pi\alpha}{f_g}}  \\ \nonumber
&\approx&   4\times 10^{-3} \eqfrac{\alpha}{0.1}^{1/2}    \eqfrac{T}{40~\rm{K}}^{5/4} 
\eqfrac{r}{70~\rm{AU}}^{5/4}  \\ \nonumber
&&\left(\frac{M_*}{1.5~M_\Sun}\right)^{-5/4}.
\end{eqnarray}
Gap opening requires  $\Delta > R_{\rm H}$ and $R_{\rm H} > H$.  The latter 
requirement is automatically satisfied for fragments formed by GI.

While the effects of GI are often parameterized through an $\alpha$ viscosity, 
\cite{BP99} have pointed out that $\alpha$, a purely local quantity, may not adequately 
describe GI driven transport, which is inherently non-local. \cite{LodRi05} have 
shown that for sufficiently thin disks, the approximation is reasonable: in order to 
form objects of planetary mass, the disk must be relatively thin and at least 
marginally within this limit. However, even in this thin-disk limit, it is not clear that GI
driven torques will exactly mimic viscous ones at gap-opening scales.

Equation (\ref{eq-gapmass}) implies that the gap opening mass is less than or 
equal to the fragment mass for effective viscosities consistent with GI. We use  
$\alpha \approx 0.1$, as this is consistent with active GI  
\citep{Gam2001,LodRi05,Krumholz2007a}.  If the local disk viscosity is lower, 
 fragments will always form above the gap-opening mass.

\subsubsection{Gap-opening starvation mass}\label{sec-gapstarve}
Gap-opening slows accretion onto the planet, but does not starve it of material 
completely. Accretion rates through gaps remain uncertain for standard core 
accretion models, and numerical models are not available for accretion onto the 
distended objects formed through GI fragmentation.  Nevertheless simulations of accretion
through gaps in low viscosity disks \citep{1999ApJ...526.1001L}
demonstrate that accretion is slower through larger gaps, and this qualitative
conclusion likely remains valid as long as gaps form.

Analogous to the isolation mass, we consider a ``gap starvation mass" that is 
related to the ratio of the gap width to planet Hill radius. Rewriting \Eq{eq-gapwidth}
 we find the ratio of gap width to Hill radius is:
\begin{equation}\label{eq-gaptohill}
\frac{\Delta}{R_H} = \left(\frac{f_g q }{\pi \alpha}\frac{r^2}{H^2}\right)^{1/3} 
\end{equation}
Note that $\Delta > R_H$ recovers the canonical gap opening estimate appropriate for
Jupiter: $q > 40 \nu /( { r^2 \Omega}) $, modulo 
order unity coefficients \citep[cf.][]{2006Icar..181..587C}.
 
If we make the simplifying assumption that gap accretion terminates when the gap 
width reaches a fixed number of Hill radii, $f_S$, we can calculate a gap starvation mass.  
We expect that for a gap to truncate accretion, $f_S \gtrsim f_H = 3.5$, the width in Hill radii
used to calculate the isolation mass (\S \ref{sec-isolation}).  Terminating accretion is an 
unsolved problem for Jupiter in our own solar system, and so to provide a further constraint on $f_S$,
we refer to the numerical simulations of \cite{2009Icar..199..338L} (See also \cite{2003ApJ...586..540D} 
for a detailed explanation of the numerical work).  Their runs 2l and 2lJ exhibit asymptotic
mass growth after $\sim$2.5 Myr for a constant-mass, low viscosity ($\alpha = 4\times 10^{-4}$) disk under 
conditions appropriate to the formation of Jupiter.  Using \Eq{eq-gaptohill}, we solve for the width of 
the gap generating this fall-off in accretion rate and find $f_S = \Delta/R_H \sim 5$.
The need for an extremely large and well-cleared gap reflects the integrated effects of low-level accretion
through the gap and onto the planet over the disk lifetime of a few Myr. Even a slow
trickle of material onto the planet can contribute to significant growth.

\begin{figure}[htbp]
   \centering
   \includegraphics[scale=0.45]{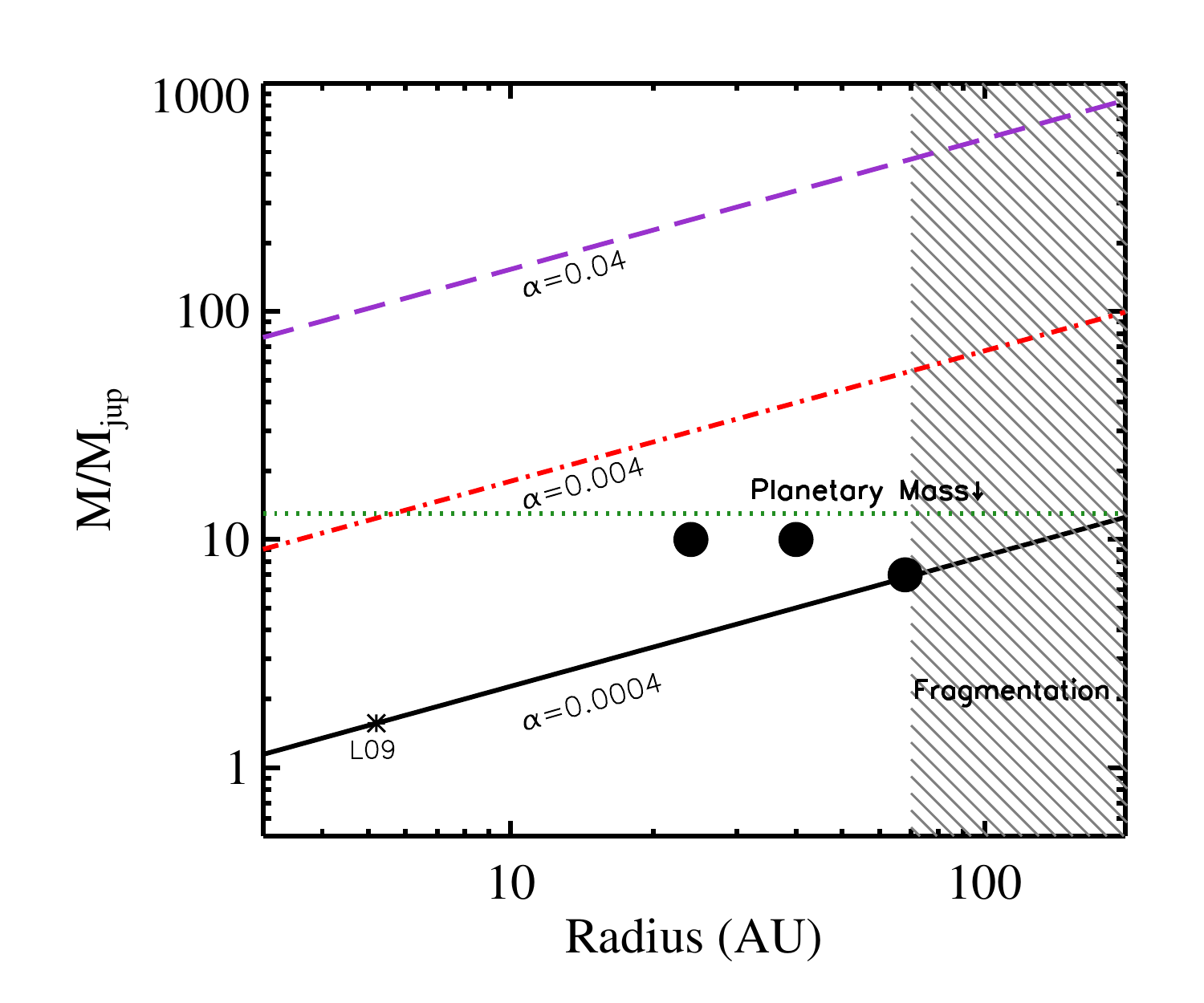} 
   \caption{The gap starvation mass as a function of disk radius.
   We show curves for several values for $\alpha$, and indicate the planetary mass 
regime, and the region in which disk fragmentation is likely. 
We use $f_S = 5$, scaled to   simulation 2lJ  of Jupiter formation in 
\cite{2009Icar..199..338L} (labeled L09 in the figure). The radial scaling 
is derived assuming $H/r \propto r^{2/7}$.  For the low viscosity case, we 
normalize the scale height to Jupiter at $5.2$ AU in a 115 K disk for comparison with L09. 
 For the higher viscosities, we normalize the disk scale height to the lowest expected 
temperatures (equation \ref{eq-minT}). For comparison we show the HR 8799 planets as black circles.}
   \label{jupiterscale}
\end{figure}

Using $f_S =5$, the gap-opening starvation mass for HR 8799 
scaled to both the simulated solar-system viscosity and to the expected GI viscosity is:
\begin{eqnarray}
M_{\rm starve} &\approx& 8 \Mjup \eqfrac{\alpha}{4 \times 10^{-4}} \eqfrac{\Delta}{5 R_H}^{3}  \label{eq-low_Tgap} \\
 &&\eqfrac{T}{40 ~\rm{K}} \eqfrac{r}{70~\rm{AU}} \nonumber \\ 
&\approx& 2000\Mjup \eqfrac{\alpha}{0.1}  \eqfrac{\Delta}{5 R_H}^{3}  \nonumber \\
&&\eqfrac{T}{40 ~\rm{K}} \eqfrac{r}{70~\rm{AU}}. \nonumber  
\end{eqnarray}
In order to limit growth to planetary masses, the effective viscosity must be two 
orders of magnitude below that expected in GI unstable disks, roughly $\alpha \sim 
10^{-3}$. More restrictively this requires that other local transport mechanisms such 
as the MRI be weaker than currently predicted by simulations---they produce
 $\alpha \sim10^{-2}$ at least in disks with a net magnetic flux 
 \citep{Flemetal00,2007A&A...476.1123F,2009ApJ...697.1269J}. \Fig{jupiterscale} illustrates the 
 scaling of gap starvation mass with radius for several values of $\alpha$.
It appears that active disks face severe obstacles in producing planetary mass objects
unless the disk disappears promptly after their formation. Although we 
expect fragmentation to make the disk more stable by lowering the local column density, 
there is little reason to expect a recently massive disk to be so quiescent.

\subsubsection{Planet starvation through gap overlap}
Although it is unlikely that the disk will fragment into a sufficiently large number of planetary 
mass objects to completely deplete the disk of mass \citep{2009MNRAS.392..413S}, the
formation of multiple fragments simultaneously may limit accretion through 
competition for disk material by opening overlapping gaps. The current separation 
between the planets is such that gaps larger than roughly three Hill radii overlap, so 
depending on their migration history, this could limit growth (see also \S\ref{sec-migration}). 
From \Eq{eq-low_Tgap} we see that if gaps are forced to be smaller by a factor of two due 
to competition with another planet, the expected masses are  decreased by a factor of 8. 
This effect would imply that fragments in multiple  systems should be lower in mass.
 Note that when simulated disks fragment into multiple objects simultaneously, they
generally have orbital configurations like hierarchical multiples rather than planetary  
systems \citep{2009MNRAS.392..413S,KMKK10}.  Whether the same conditions required 
to limit fragment growth---reduced disk mass and/or viscosity after fragmentation---can allow
the retention of a planetary system of fragments remains to be simulated.

\subsection{Disk dispersal as a means to limit fragment growth}\label{sec-growthsummary}
Mechanisms for gas dispersal such as photoevaporation may be necessary to stunt 
planetary growth, even for models of Jupiter in our own solar system 
\citep{2009Icar..199..338L}.  Dissipation timescales for A-star gas disks are 
thought to be short, less than 2-3 Myr \citep{2006ApJ...651L..49C}, which could
 halt growth before the gap-opening starvation mass is achieved.
Radiative transfer models such as \cite{2009ApJ...690.1539G} have calculated 
that photoevaporation by the central star will become important at radii of 
$\sim 100$ AU around one Myr for an A star \cite[see also][]{Ercolano09}. This timescale coincides with the 
expected fragmentation epoch, and may aid in shutting off accretion both onto the 
disk, and onto the planets.

\section{GI planet formation in the context of star formation}\label{sec-starform}
We now consider how the disk can reach the fragmentation conditions described in \S\ref{idealGI} in
the context of a model for star formation. Due to the effects of infall onto the disk, we find
that planet formation via GI can only occur  when the fragmentation epoch is concurrent
 with the end of the main accretion  phase, as the protostar transitions from a Class I to a 
 Class II object \citep{1994ApJ...420..837A}. Fragmentation at earlier times leads to the
  formation of more massive companions, while fragmentation at later times is unlikely 
  because disks are too low in mass \citep{2009ApJ...700.1502A}.
 
\subsection{Ongoing accretion and the formation of binaries and multiples} \label{sec-acctheory}
 Because the cooling time constraint is easily satisfied for expected disk temperatures, 
 disks are likely driven to fragmentation by lowering $Q$.  Even when $Q$ is above 
 the threshold for fragmentation, torques generated by self-gravity (e.g. spiral arms) 
 can drive accretion. When the infall rate onto the disk is low, a self-gravitating disk 
 regulates its surface density and hence $Q$ so that the torques are just large enough to 
 transport the supplied mass down to the star, thereby avoiding fragmentation.
 However, GI cannot process matter arbitrarily quickly because the torques saturate. 
 Thus the disk will be driven toward fragmentation if the infall rate becomes too high. 
 This critical accretion rate is a function of disk temperature: 
 \begin{equation}\label{eqn-critMdot}
\Mdot_{\rm crit} \approx\frac{ 3c_s^3 \alpha_{\rm sat} }{GQ}
\end{equation}
\citep{Gam2001,ML2005}.  Numerical simulations show that GI saturates at
$\alpha_{\rm sat} \sim 0.3-1 $ 
\citep{Gam2001,Krumholz2007a,LodRi05,KMKK10}.  
If the infall rate onto the disk exceeds $\Mdot_{\rm crit}$ the disk can no longer regulate
the surface density to keep $Q$ just above unity, and fragmentation will occur.
Because the conversion of accretion energy to thermal energy at large radii is inefficient, 
disks cannot restabilize through heating to arbitrarily high accretion rates \citep{KM06}.

If fragmentation occurs due to rapid accretion, as described above, it is difficult to limit subsequent fragment growth.
As demonstrated in \S\ref{sec-gapstarve},  gap opening does not limit accretion efficiently when the effective viscosity,
 in this case, $\alpha_{\rm sat}$, is high. 
  
A more general barrier to making small fragments during rapid infall is the large
 reservoir of material passing by the fragment as star formation proceeds 
 \citep{1994MNRAS.269L..45B}.  The  specific angular momentum, $j$, of accreting 
 disk material typically increases with time (modulo small random fluctuations in a 
turbulent core), landing at a circularization radius, $r_{\rm circ} = j^2/GM_*$,  
which is larger than the fragment's orbit. Since the fragment's Hill radius is roughly 10\% of the 
disk radius, newly accreted material undergoes many orbits in the fragment's sphere of 
influence as it tries to accrete onto the central star, and some fraction will accrete onto the
fragment itself. This process is less efficient if global GI modes drive fragments to smaller radii,
because their Hill radii shrink.  However, growth to stellar (or sub-stellar) masses may still occur 
 as long as migration timescales are not faster than accretion timescales. The latter scenario implies
 that the disk cannot fragment at early times and still reproduce a single-star system like HR 8799.
 
The trend toward continued growth and even mass equalization following disk fragmentation 
is observed in numerous simulations with ongoing accretion 
\citep{1994MNRAS.269L..45B, 2000MNRAS.314...33B,2003ApJ...595..913M,
2009MNRAS.400...13W,KMKK10,2009SciKrum}. 
Simulations of planet formation by GI with ongoing accretion also illustrate this 
behavior \citep{2009ApJ...695L..53B}.  

Consequently, if the HR 8799 planets formed by GI, they must have fragmented at the tail
end of accretion from the protostellar core onto the disk.  Most likely, this requires that the protostellar
core have properties such that its infall rate reaches the critical value in equation (\ref{eqn-critMdot}) just
as the core is drained of material.  If this coincidence in timing does not occur, fragmentation produces
a substellar, rather than a planetary, companion. Whether fragmentation at this epoch can produce 
 non-heirarchical orbits like HR 8799 remains to be explored.
 
\subsection{Reaching Instability in the absence of infall: FU Orionis outbursts}\label{sec-fuori}
Driving the disk unstable with external accretion corresponds to excessive 
growth of fragments. It is therefore tempting to consider mechanisms to lower $Q$ 
through disk cooling, while holding the column density fixed.  Because the disks 
are dominated by irradiation, changes in the viscous dissipation due to accretion 
are unlikely to affect a significant temperature change, and so lowering $Q$ requires 
order of magnitude changes in the stellar luminosity due to the weak scaling of 
$T \propto L^{2/7}$.   FU Orionis type outbursts \citep{1996ARA&A..34..207H} 
can cause rapid changes in luminosity.  To result in planetary mass fragments, the
 luminosity drop following an FU Ori outburst would need to reach at least the minimum
 luminosity used in equation (\ref{eq-minT}) on timescales shorter than an outer disk 
 dynamical time.

The accretion of GI formed embryos onto the protostar is a proposed source of the 
outbursts  \citep{VorBas06}. If this occurs, perhaps a final generation of 
gravitational fragments, the so-called ``last of the Mohicans" 
\citep{1997MNRAS.285..403G}, would remain as detectable companions. 
Although the lack of infall in this scenario might ease the gap opening constraints, 
fragments face all of the other difficulties discussed above in remaining low in mass.

\section{Migration in a multi-planet system}\label{sec-migration}
A final consideration for making wide orbit planets through GI is their subsequent 
migration history.  If formed via GI, the HR 8799 planets had to migrate inward to 
their current locations.  There is an independent reason to believe that migration did
 in fact take place in this system. As discussed by \cite{FabMur09}, the long term stability of 
HR 8799 requires a resonant orbital configuration, most plausibly a 4:2:1 mean motion 
resonance, which likely resulted from convergent  migration in the protoplanetary gas disk.
We demonstrate below that while this history is plausible, it requires special disk conditions.

Inward Type II migration  (appropriate for gap opening planets) is expected at formation, 
because the fragmenting region is within the part of the disk accreting onto the star. 
 Type II migration of a single planet occurs on the disk viscous timescale:
 \begin{eqnarray}\label{eq-visctime}
\tau_\nu &\approx&{\frac{r^2}{\nu}}  =\frac{ r^2 \Omega }{\alpha c_s^2} \\ 
&\approx & 0.4\,\rm{Myr} \eqfrac{r}{70 ~\rm{AU}}^{13/14}\eqfrac{M_*}{1.5\Msun}^{1/2} \eqfrac{\alpha}{0.1}^{-1}
\end{eqnarray}
where we have used  $\nu = \alpha c_s^2/ \Omega$.  This timescale is short enough to allow substantial
migration during the lifetime of the gas disk.

If  fragments migrated inward independently, they could never become captured 
in resonance, as the innermost planet would migrate farther and farther from its 
neighbor. However, as shown in \S \ref{sec-GIaccrete}, planetary gaps may overlap 
if they are within several Hill radii of each other, comparable to the current separations 
between the HR 8799 planets.  This overlap alters the torques felt by, and thus the migration
 of, the planets.  As shown by \cite{2000MNRAS.313L..47K}, if multiple planetary gaps
 interact, convergent migration is possible. Gap interaction allows  an outer planet to shield 
 an inner planet from the material, and thus torques, of the outer disk, slowing or halting its 
 inward migration and allowing the outer planet to catch up. This mechanism is invoked
  by \cite{2002ApJ...567..596L} to generate convergent migration and resonance capture 
  in the planets orbiting GJ 876.

Under the assumption that gap overlap allows convergent migration and resonance capture, 
we now ask:  What is the overall direction of the subsequent migration?  
We note that if the planets migrated a substantial distance after resonance capture, 
eccentricity damping by the gas disk was likely necessary \citep[c.f.,][]{2002ApJ...567..596L}.  
We do not consider eccentricity damping further here. Once two planets are caught in 
mean-motion resonance, the torque on an individual planet from the gas disk can cause
 both planets to migrate, with angular momentum transfer mediated by the resonance.  
 \cite{2001MNRAS.320L..55M} \citep[see also][]{crida09} have argued that the torque 
 imbalance on a pair of gap-opening resonant planets can even reverse the direction of migration,
 although this relies on a significant difference between planet masses. 

Nevertheless, understanding the planets' overall migration requires understanding how 
overlapping gaps alter the torque balance on the group of planets. 
Guided by our interest in clean gaps that limit the growth of planets
 (\S\ref{sec-GIaccrete}--\ref{sec-starform}), we consider the following simplified problem.   

We imagine that the three planets have cleared, and are embedded in, a single large gap 
which is sufficiently clean that any disk gas passing through is dynamically unimportant.  
Because the system is locked in a double mean-motion resonance, we assume that an 
imbalance in the torques acting on the two edges of the gap can cause all three planets to migrate.
We can then ask:  is there a sufficient flux of angular momentum from the outer disk to cause the 
planets to migrate inward with the viscous accretion of the disk? 

When in the 4:2:1 resonance, the total angular momentum of the planets is roughly
 $ 2.4M_{p} \Omega_p r_p^2$, where we have assumed that the planets are roughly
  equal in mass, $M_p$, while $r_p$ and $\Omega_p$ are the separation and Keplerian 
  angular velocity of the outermost planet. The angular momentum flux from the outer disk 
  is large enough to move the planets on a viscous time $r_p^2/\nu$ when:
\begin{equation}
\Mdot \Omega_p r_p^2 \gtrsim 2.4M_{p} \Omega_p r_p^2 (\nu/r_p^2)  \;,
\end{equation}
where $\dot M = 3\pi\Sigma\nu$ is the mass flux through radius $r_p$.

Using the above inequality, we can calculate a critical disk surface density at the 
current location of the outer planet such that the disk can push the planets inward:
\begin{equation}
\Sigma \gtrsim \frac{M_p}{4 r_p^2}.
\end{equation}
For $M_p = 10\Mjup$ and our fiducial disk temperatures (equation \ref{eq-minT}), this 
constraint is always satisfied when $Q=1$.  The disk is unable to  cause inward migration 
when $\Sigma \lesssim 4\rm{g/cm^2}$ at 70 AU, which is equivalent to $Q\sim20$.  

If the planets do share a clean common gap, a large fraction of the disk would be effectively
 cleared of gas while a massive outer disk is still present.  A similar mechanism has been invoked 
 to explain transitional disks, which contain holes at radii of a few tens of AU and smaller
 \citep{2002ApJ...568.1008C}. Transport of disk gas through a less well-cleared gap could 
  substantially alter this picture.

In summary, it is possible to envision a scenario in which the HR 8799 planets 
migrate inward to their current locations in such a way that their orbits
converge, allowing resonance capture. This scenario is consistent with other 
constraints on GI planet formation: shortly after formation, the disk must have low 
accretion rates and decline in mass in order to (a) limit the growth of fragments, (b) allow for large, overlapping gaps.    

 More stringent constraints will require future work
 on migration in gravitationally unstable disks, particularly in the presence of multiple planets 
 massive enough to clear large, overlapping gaps.

\begin{figure*}[htbp]
   \centering
   \includegraphics[scale=0.3]{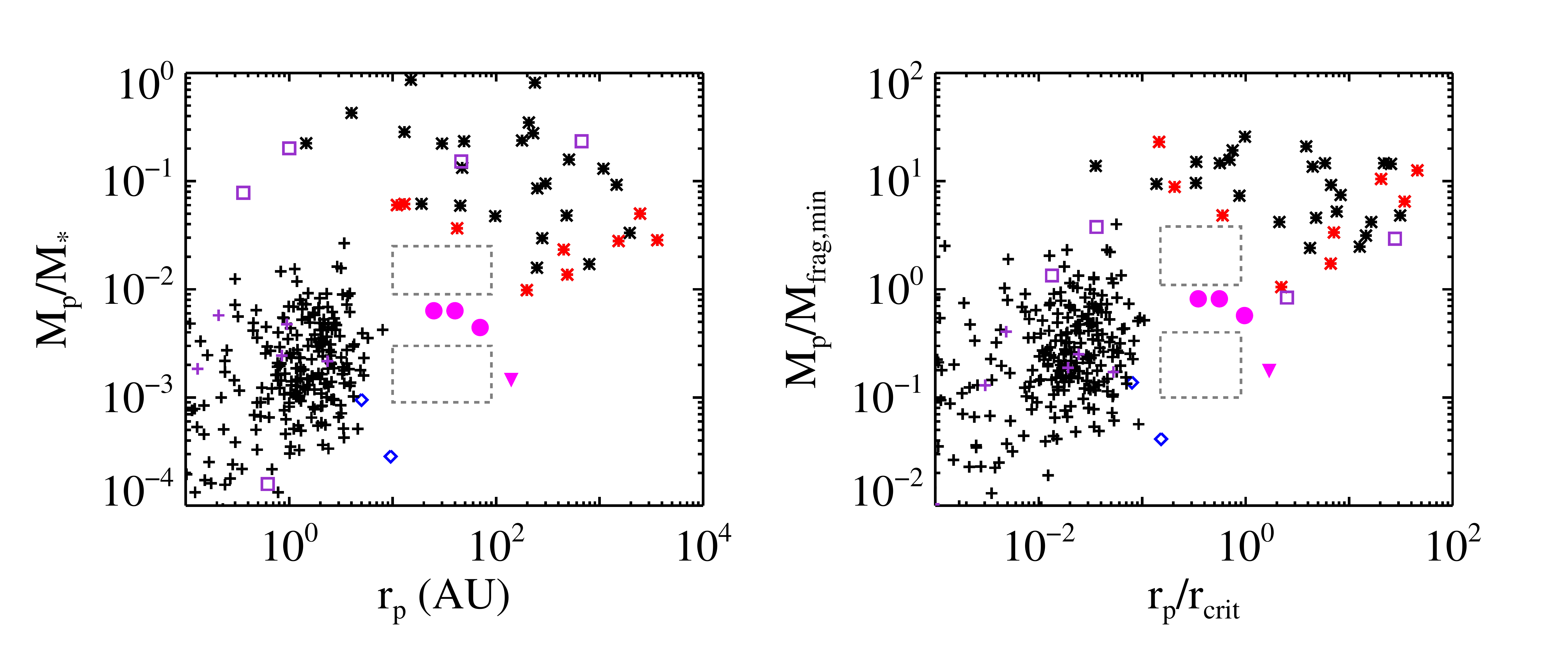} 
\caption{(Left) Known substellar companions (stars) and planets (plusses) as a function of mass ratio and projected 
separation.  The three objects in the HR  8799 system are shown by pink circles, and a pink triangle denotes the 
upper-limit mass ratio for Fomalhaut b based on the dynamical mass estimate of \cite{2009ApJ...693..734C}.   
Grey squares indicate the gap regions. Ongoing surveys are necessary to determine whether there is a continuous
 distribution between Jupiter/Saturn (blue diamonds) and HR 8799, or HR 8799 and brown dwarf companions. 
Planets around very low mass primaries with $M_* = 0.02$--0.1$\Msun$ are marked by purple squares.  
These systems have mass ratios more akin to the substellar companions than to the remainder of the population of planets.  
Primary masses range from $M_* = 0.02\Msun$--$1\Msun$ (black) and $M_* = 1\Msun$--2.9$\Msun$ (red) for substellar
 companions and from $M_* = 0.1$--0.4$\Msun$ (purple) and $M_* = 0.4$--4.5$\Msun$ (black) for planets.
(Right) The same objects plotted as function of the minimum fragment mass, $M_{\rm frag,min}$ and
critical radius, $r_{\rm crit}$.  We use \Eq{eq-rminthick} to calculate $r_{\rm crit}$.  For $M_{\rm frag,min}$, 
we apply \Eq{eq-mfragt} at radius $r_{\rm crit}$ under the simplified assumption that the disk temperature is 
set by the stellar luminosity: $L/L_\odot =  (M_*/\Msun)^{3.5}$ for $M_* > 0.43 \Msun$ and $L \propto M_*^{2.3}$
for lower-mass stars.  The temperatures used to calculate fragment masses are not allowed to dip below 20K.  
Masses below $M_{\rm frag,min}$ are unlikely to result from GI.}
   \label{fig-bdplot}
\end{figure*}

\section{Current Observational Constraints}\label{sec-observe}
While there is a regime of parameter space in which planet formation is possible, 
typical conditions produce more massive ($> 13 \Mjup$ ) companions.  If GI 
fragmentation ever forms planets, then fragmentation should typically form more 
massive objects. Consequently, these planets would constitute the low mass 
tail of a distribution of  disk-born companions. If the mass distribution is continuous 
there should be more sub-stellar companions than planets at comparable distances of 50-150 
AU. Observing this population is a strong constraint on the formation mechanism,
but current data are insufficient to draw conclusions. 

\cite{2009A&A...493.1149Z} have compiled the known sub-stellar companions in 
this range of radii to date. This range of separations falls beyond  the well-established
 inner  brown dwarf ``desert" \citep{2004AJ....127.2871M}, and has not been well probed
  due to observational difficulties at these low mass ratios. Note that the overall dearth of  
  brown dwarf companions to solar mass stars is not a selection effect
\citep{2009ApJS..181...62M,2009A&A...493.1149Z}.  

 We illustrate the observational constraints by plotting the companions from 
 \cite{2009A&A...493.1149Z}  along with  the known exoplanets compiled by the
 Exoplanet Encyclopedia\footnote{October 2009, http://exoplanet.eu, compiled
 by Jean Schneider} as a function of mass ratio and projected separation in \Fig{fig-bdplot}a,
 and as a function of minimum fragment masses and fragmentation radii in \Fig{fig-bdplot}b. 
 We compare these with the HR 8799 planets, Fomalhaut b, and the solar system giants. We 
 distinguish between stars of different masses because disk fragmentation becomes more likely for higher 
 mass stars \citep{KMK08}.
 
 At present, neither the population of substellar companions nor the population of exoplanets is 
 continuous out to HR 8799. Many selection biases are reflected in \Fig{fig-bdplot}, and these gaps in
 particular may be due to  selection effects; resolution and sensitivity make it difficult to detect both
 wide orbit planets, and close-in low mass brown dwarfs. We note that while there are actually fewer
planets at distances less than 1 AU, the cutoff above 5 AU is unphysical. There is not yet any 
indication of an outer cut off radius in the exoplanets: if they  continued out to larger separation, 
the distribution would easily encompass the HR 8799 and Fomalhaut systems.

 Data from ongoing surveys like that which found HR 8799 are necessary to verify the true companion distribution as
a function of mass and radius. If  these planets are formed through GI, we would expect observations to fill in the gap 
between HR 8799 and higher mass ratio objects to show a continuous distribution. 
If these planets are formed via core accretion, than observations may fill in the plot on
the opposite side of HR 8799, occupying a region of parameter space for which neither 
core accretion nor GI is currently a successful formation mechanism. 

\section{Summary} \label{sec-discussion}

We have demonstrated that while GI-driven fragmentation is possible at wide 
distances from A stars, fragment masses typically exceed the deuterium burning 
``planet" limit. In contrast, the formation of sub-stellar and stellar companions is 
more likely because moderate disk temperatures and active accretion onto and 
through the disk drive disk-born objects to higher masses.

If the HR 8799 planets did form by GI, the following criteria had to be met:
\begin{enumerate}
\item{Fragments should form beyond $40- 70$ AU: inside of this location the disk 
will not fragment into planetary mass objects even if $ Q\lesssim1$. Grain growth is required for
fragmentation at the lower end of this range.}
\item{Temperatures must be colder than those of typical disks to limit the initial fragment masses.}
\item{The disk must be driven unstable at a special time: infall onto the disk must 
be low, but the disk must remain massive (e.g. the end of the Class I phase). The 
disk must only become unstable to fragmentation at this point because earlier 
episodes of instability should lead to sub-stellar or stellar companion formation.}
\item{The subsequent growth of fragments must be limited through efficient gap 
clearing necessitating low disk viscosity or early gap overlap. Disk dispersal via 
photoevaporation may also be necessary.}
\item{The three fragments must form at the same epoch separated by several Hill 
radii, implying that the entire outer disk becomes unstable simultaneously.}
\item{Migration must be convergent. This likely requires the gaps of the planets to 
overlap so as to starve the inner most planet of disk material, thereby preventing 
runaway inward migration.}
\end{enumerate}

If these conditions are met, then the planets in HR 8799 could comprise the low-mass
 tail of the disk-born binary distribution, the runts of the litter. In this case  one 
would expect to find a larger number of brown dwarfs or even M stars in the same 
regime of parameter space -- surrounding A-stars at distances of $50-150$ AU.  

Ongoing direct imaging surveys of A and F stars will provide a strong constraint on 
the formation mechanism for this system: if HR 8799 is the most massive of a new 
distribution of widely separated planets, our analysis suggests that formation by GI is unlikely. On the 
contrary, the discovery of a population of brown dwarf and M-star companions to 
A-stars would corroborate formation via disk fragmentation.

\begin{acknowledgements}
{\bf Acknowledgements: }
The authors thank Eric Ford and Jonathan Tan for organizing the 
Astrowin workshop at the University of Florida, which inspired this research, Chris 
Matzner, for invaluable comments on this manuscript, and Yoram Lithwick, 
Roman Rafikov, Ken Rice,  Pawel Artymowicz, Eugene Chiang, 
Sally Dodson-Robinson, Aaron Boley, and Stella Offner for helpful 
discussions. The authors also thank an anonymous referee for
helpful comments. KMK is supported in part by an Ontario Graduate Scholarship. RMC 
is funded by an  Institute for Theory and Computation Fellowship at Harvard 
University.
\end{acknowledgements}

\appendix

\section{A. Cooling and Fragmentation in Irradiated disks}\label{sec-irradcool}

\cite{RR09} has suggested that the cooling time might be altered in an irradiated disk. 
Here we consider the cooling time for thermal perturbations to a disk, and show 
that a simple formula [\Eq{eq-t_lum}] gives the cooling time for arbitrary levels of 
irradiation, at least in radiative optically thick disks.
 
Consider ambient radiation striking an optically thick disk with a normal flux
$F_{o} = \sigma T_o^4 \equiv (3/8) F_{\rm irr}$ (where the numerical factor in the last definition is purely for later convenience).  
Note that $T_o$ depends both on the irradiation field (from the host star and/or 
light emitted and reflected from a surrounding envelope) and also on the disk's 
surface geometry, e.g.\ flaring angle.  
By incorporating these variables into $T_o$ we attempt a general calculation.
At the photosphere, where the optical 
depth to the disk's self-emission $\tau = \tau_{\rm phot} \approx 1$, energy balance gives:
\begin{equation} \label{eq-eb}
\sigma T_{\rm eff}^4 \simeq \sigma T_o^4 + F/2  \, .
\end{equation} 
Here, $F$ is the luminous flux from any internal sources of energy dissipation, 
e.g.\ viscous accretion, shocks, or the gravitational binding energy released by a 
collapsing fragment.  The factor of two reflects that half of the radiation is emitted 
from each surface of the disk.  
Since optical light is absorbed above the IR photosphere, about half the 
irradiation free streams out before it heats the disk \citep{1997ApJ...490..368C}.  
We can absorb this reduction into the definition of $T_o$. 

We assume heat is transferred by radiative diffusion as:
\begin{equation}
{4 \over 3} \sigma  {d T^4 \over d\tau} =  {F \over 2}.
\end{equation}
since convection is suppressed by irradiation and may be a negligible correction in any event \citep{RR07}.
Integration from the midplane at $\tau = \tau_{\rm tot}$  and $T = T_{\rm m}$ to the 
photosphere gives:
\begin{equation} \label{eq-tauF}
(4/3)\sigma (T_{\rm m}^4 - T_{\rm eff}^4) = (F/2)( \tau_{\rm tot} - \tau_{\rm phot})\, .
\end{equation} 
We now drop the subscripts from $\tau_{\rm tot}$ and the midplane $T_{\rm m}$, 
which we will soon take as the characteristic temperature (ignoring order unity 
corrections from height averaging).  Furthermore we apply \Eq{eq-eb} and take the $\tau_{\rm phot} \ll \tau_{\rm tot} \rightarrow \tau$ limit (meaning that we don't need to know the precise location of the photosphere) to express 
\begin{equation} \label{eq-F}
F \simeq {8 \sigma \over 3 \tau}(T^4 - T_o^4) = {1 \over \tau} \left({8 \sigma T^4 \over 3 }
-F_{\rm irr} \right)\, ,
\end{equation} 
which shows that the midplane temperature is controlled by the larger of $F \tau$ and $F_{\rm irr}$.  

The cooling timescale to radiate away thermal fluctuations (generated e.g.\ by GI) is
\begin{equation} \label{eq-t_lumdef}
t_{\rm cool} = {\Sigma \delta U \over \delta F} 
\end{equation} 
 where a temperature perturbation $\delta T$ has an excess heat $\delta U 
\approx c_P \delta T$, and $c_P = (k/\mu)\gamma/(\gamma-1)$ is the specific heat 
for a mean molecular weight $\mu$ and adiabatic index $\gamma$.  
Strongly compressive motions, which are not at constant pressure,
will introduce order unity corrections that we ignore. 

The excess luminous flux, using \Eq{eq-F}, is
\begin{eqnarray} \label{eq-deltaF}
\delta F 
&=& {8 \sigma \over 3 \tau}{\delta T \over T} \left[ (4-\beta) T^4 + \beta T_o^4\right]  \nonumber \\
&=&{32 \sigma T^3 \delta T \over  3 \tau} \times \left\{
\begin{array}{cl}
(1 - \beta/4) &~{\rm if}~ T \gg T_o, \beta \neq 4 \\
1 &~{\rm if} ~T \simeq T_o
\end{array}
 \right.\, ,
\end{eqnarray} 
where $\tau = \kappa \Sigma/2 \propto T^\beta$.  The point is that the escaping 
flux varies by only an order unity factor between the strongly ($T \simeq 
T_o$) and weakly ($T \gg T_o$) irradiated regimes.  Typical grain opacities, $0 < 
\beta < 2$, ensure the correction is order unity (and also ensure that we can 
ignore the catastrophic heating that would occur if $\beta > 4$).

Combining \Eqs{eq-t_lumdef}{eq-deltaF} with the definition of heat capacity we 
find that the cooling time is simply
\begin{equation} \label{eq-t_lum}
t_{\rm cool}  \approx {3 \gamma \Sigma c_{s}^2 \tau \over 32  (\gamma -1) \sigma T^4} \times \left\{
\begin{array}{cl}
(1 - \beta/4) ^{-1}&~{\rm if}~ T \gg T_o, \beta \neq 4 \\
1 &~{\rm if} ~T \simeq T_o
\end{array}
 \right.\, ,
\end{equation} 
where the isothermal sound speed $c_{s} = \sqrt{kT/\mu}$.

From this derivation, we see that the cooling time obeys the simple form of 
\Eq{eq-t_lum}  for all levels of irradiation --- which only introduces an order unity  
$\beta$ correction.  
The cooling time depends on $\beta$ for weakly irradiated disks because 
changes to the opacity alter the amount of flux that escapes from the midplane.
In highly irradiated disks, opacity changes have little effect because 
the small difference between the midplane and surface temperatures drives a 
weak flux.

While increasing the irradiative flux incident on a disk decreases the cooling 
time by raising $T$, it will not trigger fragmentation in a  $Q \sim 1$ disk, since it 
increases $Q$.
We will not explore optically thin or convective disks at this time.

We assume that $\Omega t_{\rm cool} < \zeta \sim 3$ is the fragmentation criterion 
independent of irradiation.  When the cooling time is longer, the disk is presumed 
to enter a state of gravito-turbulence \citep{Gam2001}. We take this term to mean 
a quasi-steady state of gravitationally 
driven turbulence, on scales  $\lesssim H$ wherein viscous dissipation of GI turbulence
regulates $Q \sim 1$. 
Thus the cooling time can be translated to a the critical value of $\alpha$ 
at which gravito-turbulent accretion disks will fragment.  While \citet{Rice05} find 
that an $\alpha$-threshold is more robust than one for $t_{\rm cool}$ when the 
adiabatic index varies, they did not include irradiation, which we contend would 
reveal that cooling is ultimately the more physical criterion, but the issue is best 
settled by simulation.
The emitted flux with an $\alpha$-viscosity and $Q \sim 1$ gives
\begin{equation} 
F \approx (9/4) \nu \Sigma \Omega^2 \approx 9/(4\pi) \alpha c_{s}^3 \Omega^2/G
\end{equation} 
which combined with \Eq{eq-F} and  \Eq{eq-t_lum} without the 
$\beta$ correction gives
\begin{equation} 
\Omega t_{\rm cool} \approx \left({\gamma \over \gamma -1} \right){\Omega \Sigma c_{s}^2 \over 4\left(F +{ F_{\rm irr} / \tau}\right)} \approx \left({\gamma \over \gamma -1}\right) {1 \over 9 \alpha}\left(1 +{ F_{\rm irr} \over F \tau}\right)^{-1} 
\end{equation} 
When irradiation is weak enough that $F_{\rm irr} \lesssim F \tau$, we recover the 
standard $\alpha \gtrsim 1$ criterion for fragmentation (ignoring the accumulated 
order unity coefficients).  However for stronger irradiation with 
$F_{\rm irr} \gtrsim F \tau$, fragmentation occurs for  $\alpha \gtrsim 
F \tau/F_{\rm irr}$, a lower threshold. 

Gravito-turbulent models require modification when $F_{\rm irr} \gtrsim F$, i.e.\ an 
even lower level of irradiation than needed to affect the $\alpha$ fragmentation 
threshold.  In this case the disk shows some similarities to isothermal disks,
and should have lower amplitude density perturbations, because there is 
insufficient viscous dissipation to support order unity thermal perturbations (see 
the related discussion in \citealp{RR09}).  We note that simulations of isothermal 
disks do develop GI, exhibit GI-driven transport and fragment 
\citep{Krumholz2007a,KMKK10}, but they do not appear particularly turbulent.

\section{B. Temperature due to Viscous Heating} \label{sec-visc}
We now show that viscous heating is relatively unimportant in the outer reaches of irradiated A-star disks
(see also section 3 of \citealt{RR09}). For an optically thick disk with an ISM opacity law, balancing viscous heating and emitted radiation gives
\begin{equation}
{8 \over 3 \tau} \sigma T^4 \approx \frac{3}{8\pi} \dot{M} \varOmega^2
\end{equation} 
The solution for the midplane temperature is 
\begin{equation} 
T \approx \left({9 \over 128 \pi^2} {\dot{M}\kappa_o \sqrt{k/\mu} \over \sigma G Q_o}
\right)^{2/3} \varOmega^2 \approx  9 ~{\rm K} \eqfrac{\Mdot}{10^{-6}\Msun/\rm{yr}}^{2/3} 
Q_o^{-2/3} \eqfrac{r}{70 ~\rm{AU}}^{-3},
\end{equation} 
lower than the irradiation temperatures shown in \Fig{fig-cartoon} by more than a factor of four.
Note that the surface density falloff
\begin{equation} 
\varSigma = {c_s \varOmega \over \pi G Q_o} \approx 35 ~{\rm g/cm^2}\eqfrac{r}{70~\rm{AU}}^{-3}\eqfrac{\Mdot}{10^{-6}\Msun/\rm{yr}}^{1/3} \, 
\end{equation} 
is also quite steep for a constant $Q_0 =1$, viscous disk with the ISM 
opacity law.

If the disk is optically thin, then the balance between heating and cooling gives:
\begin{eqnarray} \label{eq-tthin}
 4 \tau \sigma T^4 &\approx& \frac{3}{8\pi} \dot{M} \varOmega^2\\
 T &\approx& \left({3 G \dot{M} Q_o \varOmega \over 16 \sigma \kappa_o \sqrt{k/\mu}}\right)^{2/13}  \\
& \approx& 9~{\rm K}     \eqfrac{\Mdot}{10^{-6}\Msun/\rm{yr}}^{2/13}\eqfrac{M_*}{1.5\Msun}^{1/13}   Q_o^{2/13} \eqfrac{r}{70~\rm{AU}}^{-3/13} 
\end{eqnarray} 
This temperature profile is shallow, but still colder than the irradiation temperature at large radii where the disk becomes optically thin.

 \end{document}